\documentclass{article}[10pt]
\usepackage{spconf,amsmath,epsfig}
\usepackage{amsfonts}
\usepackage{times,amsmath,epsfig,psfrag,amsfonts}
\usepackage[usenames]{color}
\usepackage[dvips]{pstcol}

\title{Set-Membership Conjugate Gradient Constrained Adaptive Filtering Algorithm for Beamforming}
\name{Lei Wang and Rodrigo C. de Lamare }
\address{Communications Research Group, \\
Department of Electronics\\
University of York, York YO10 5DD, UK\\
lw517@ohm.york.ac.uk;~rcdl500@ohm.york.ac.uk}

\begin{document}
\maketitle
\begin{abstract}
We introduce a new linearly constrained minimum variance (LCMV)
beamformer that combines the set-membership (SM) technique with the
conjugate gradient (CG) method, and develop a low-complexity
adaptive filtering algorithm for beamforming. The proposed
algorithm utilizes a CG-based vector and a variable forgetting factor
to perform the data-selective
updates that are controlled by a time-varying bound related to the
parameters. For the update, the CG-based
vector is calculated iteratively (one iteration per update) to
obtain the filter parameters and to avoid the matrix inversion. The
resulting iterations construct a space of feasible solutions that
satisfy the constraints of the LCMV optimization problem. The
proposed algorithm reduces the computational complexity
significantly and shows an enhanced convergence and tracking
performance over existing algorithms.\\

\end{abstract}
\section{Introduction}
Beamforming is an ubiquitous task in adaptive filtering and
array signal processing problems, and has been found widespread applications
in radar, sonar and wireless communications. Among the existing
techniques, the most promising one is the optimal
linearly constrained minimum variance (LCMV) beamformer \cite{Frost}
due to its simplicity and effectiveness. The LCMV beamformer aims to
suppress interference at the array output while improving the
reception of the desired signal. The constraint corresponds to prior
knowledge of the array response of the desired user $\boldsymbol
a(\theta_0)\in\mathbb C^{m\times1}$, where $\theta_0$ is the
direction of arrival (DOA) of the desired signal and $m$ is the number
of sensor elements in the array.

The optimal LCMV beamformer requires the computation of the inverse
of the covariance matrix $\boldsymbol R=\mathbb E[\boldsymbol
r\boldsymbol r^H]\in\mathbb C^{m\times m}$ with $\boldsymbol
r\in\mathbb C^{m\times1}$ being the received vector, and results in
a heavy computational load. Many adaptive filtering algorithms have
been reported to realize the beamformer design efficiently. The
well-known adaptive algorithms include stochastic gradient (SG),
recursive least squares (RLS), affine projection (AP), and conjugate
gradient (CG) \cite{Haykin}-\cite{Ahmad}. The SG algorithm is simple
to implement but suffers from a slow convergence rate and the
misadjustment. The RLS algorithm enjoys fast convergence but is more
complex to implement and may become unstable due to the divergence
problem and numerical problems \cite{Haykin}. The AP algorithm
requires the inversion of a matrix whose dimension is given by the
projection order, which results in a heavy computational load of the
AP algorithm if chosen as a large number. Besides, the convergence
of the AP algorithm is often much slower than the RLS. The CG
algorithm has a good tradeoff between performance and complexity
since it has a faster convergence rate than the SG and AP
algorithms, and requires a lower computational cost when compared
with the RLS algorithm. Many adaptive CG algorithms have been
reported in \cite{Mandyam}-\cite{Ahmad}, and the references therein.
Modified CG versions based on the LCMV criterion can be found in
\cite{Wang}, whereas for other subspace-based algorithms the reader
is referred to \cite{delamaresp}-\cite{barc}.

In this paper, we introduce a more economic adaptive algorithm based
on the CG method for the LCMV beamformer design. The proposed
algorithm utilizes the set-membership (SM) technique \cite{Diniz},
\cite{Huang} to enforce the constraints and to reduce the
computational complexity significantly without performance
degradation. The SM specifies a bound on the magnitude of the array
output and performs data-selective updates to estimate the filter
parameters. It involves two steps: 1) information evaluation and 2)
parameter update. If step 2) does not occur frequently, and step 1)
does not require much complexity, the overall complexity can be
saved substantially. SM algorithms based on the SG and RLS
methods have been reported in \cite{Diniz}-\cite{Lamare}. Here, we use
the SM technique in the CG algorithm that was reported in
\cite{Wang}, and develop a new adaptive algorithm, which is termed
SM-CG. Specifically, we define a new LCMV optimization problem
related to a constraint on the bound of the array output, and
perform the filter optimization to calculate the solution. A
parameter dependent time-varying bound is employed to measure the
quality of the filter parameters that could satisfy the constraints
and to improve the tracking performance in dynamic scenarios. The
parameters are only updated if the bounded constraint cannot be
satisfied. For the update, we define a new CG-based vector
$\boldsymbol v\in\mathbb C^{m\times1}$ to create a relation with
$\boldsymbol R$ and $\boldsymbol a(\theta_0)$, namely, $\boldsymbol
v=\boldsymbol R^{-1}\boldsymbol a(\theta_0)$. The proposed algorithm
calculates $\boldsymbol v$ via one iteration per update to obtain
filter parameters without the matrix inverse. The updated parameters
are encompassed in a parameter space, in which each member is
consistent with the bounded constraint and the constraint on the
array response based on the optimization problem. Compared with the
existing algorithms, the proposed algorithm exhibits an enhanced
convergence and tracking performance with relatively low
computational complexity. Simulation results illustrate
the performance of the proposed SM-CG algorithm.


\section{System Model and Beamformer Design}

Let us suppose that $q$ narrowband signals impinge on a uniform
linear array (ULA) of $m$ ($q \leq m$) sensor elements. The sources
are assumed to be in the far field with DOAs $\theta_{0}$, \ldots,
$\theta_{q-1}$. The received vector $\boldsymbol r$ can be modeled
as
\begin{equation} \label{1}
\centering {\boldsymbol r}={\boldsymbol A}({\boldsymbol
{\theta}}){\boldsymbol s}+{\boldsymbol n},
\end{equation}
where
$\boldsymbol{\theta}=[\theta_{0},\ldots,\theta_{q-1}]^{T}\in\mathbb{R}^{q
\times 1}$ is the DOAs, ${\boldsymbol A}({\boldsymbol
{\theta}})=[{\boldsymbol a}(\theta_{0}),\ldots,{\boldsymbol
a}(\theta_{q-1})]\in\mathbb{C}^{m \times q}$ composes the steering
vectors ${\boldsymbol a}(\theta_{k})=[1,e^{-2\pi
j\frac{d}{\lambda_{c}}cos{\theta_{k}}},\ldots,e^{-2\pi
j(m-1)\frac{d}{\lambda_{c}}cos{\theta_{k}}}]^{T}\in\mathbb{C}^{m
\times 1},~~~(k=0,\ldots,q-1)$, where $\lambda_{c}$ is the
wavelength and $d=\lambda_{c}/2$ is the inter-element distance of
the ULA, and to avoid mathematical ambiguities, the steering vectors
$\boldsymbol a(\theta_{k})$ are considered to be linearly
independent, ${\boldsymbol s}\in \mathbb{C}^{q\times 1}$ is the
source data, ${\boldsymbol n}\in\mathbb{C}^{m\times 1}$ is the white
Gaussian noise, and $(\cdot)^{T}$\ stands for the transpose. The
output of a narrowband beamformer is
\begin{equation} \label{2}
\centering y={\boldsymbol w}^H{\boldsymbol r},
\end{equation}
where ${\boldsymbol
w}=[w_{1},\ldots,w_{m}]^{T}\in\mathbb{C}^{m\times 1}$ is the complex
weight vector of the adaptive filter, and $(\cdot)^{H}$ stands for
the Hermitian transpose. For the optimal LCMV beamformer, the aim is to solve the
optimization problem
\begin{equation}
\begin{split}
\textrm{min} ~\mathbb E[|y|^2] & =\boldsymbol
w^H\boldsymbol R\boldsymbol w \\
{\rm subject~ to} ~~~
\boldsymbol w^H\boldsymbol a(\theta_0& )=\gamma, \label{lcmvprob}
\end{split}
\end{equation}
where $\gamma$ is a constant, and $\boldsymbol R$ is the covariance matrix of the received vector.
The solution of (\ref{lcmvprob}) is $\boldsymbol
w_{\textrm{opt}}=\frac{\gamma\boldsymbol R^{-1}\boldsymbol
a(\theta_0)}{\boldsymbol a^H(\theta_0)\boldsymbol R^{-1}\boldsymbol
a(\theta_0)}$. The SG, RLS, AP, and CG algorithms have been employed
to realize the design in different ways. Among them, the CG-based
algorithms exhibit some advantages due to their attractive tradeoff between performance and complexity.

\section{Proposed SM-CG Algorithm}

In this section, we introduce a new constrained optimization
strategy that combines the SM technique with the LCMV design and
utilizes the CG-based adaptive filtering algorithm.

\subsection{Time-varying SM-CG scheme}

\begin{figure}[h]
    \centerline{\psfig{figure=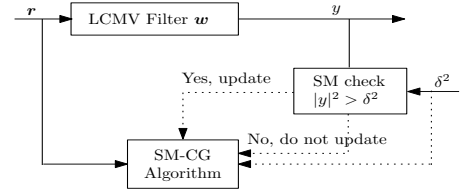,height=25mm,width=60mm} }
    \caption{Proposed SM-CG structure}
    \label{fig:model}
\end{figure}

In the proposed scheme depicted in Fig. \ref{fig:model}, the
received vector $\boldsymbol r(i)$ is processed at time instant $i$
by the LCMV filter controlled by the adaptive CG algorithm to
generate the output $y(i)$. For the existing CG algorithms
\cite{Mandyam}-\cite{Chang}, it is necessary to update ${\boldsymbol
w}(i)$ for each time instant $i$ with many iterations to obtain a
good performance. In the proposed scheme, the SM technique is
embedded to specify a time-varying bound $\delta(i)$ on the
amplitude of $y(i)$. The update only performs if the bounded
constraint $|y(i)|^2\leq|\delta(i)|^2$ cannot be satisfied. For each
update, some valid estimates of ${\boldsymbol w}(i)$ satisfy the
bound. Thus, the solution to the proposed scheme is a set in the
parameter space.

The time-varying bound $\delta(i)$ is a unique coefficient for the
proposed scheme to check if the update is carried out or not. It is
better if $\delta(i)$ could reflect the characteristics of the
environment since it benefits the tracking of the proposed
algorithm. We introduce a parameter dependent bound (PDB) that is
similar to the work in \cite{Guo} and involves the evolution of
${\boldsymbol w}(i)$ obtained from the proposed algorithm. The
time-varying bound is
\begin{equation}\label{3}
\delta(i)=\beta\delta(i-1)+(1-\beta)\sqrt{\alpha\|\boldsymbol
w(i-1)\|^2\hat{\sigma}_n^2(i)},
\end{equation}
where $\beta$ is a forgetting factor that should be set to guarantee
a proper time-averaged estimate of the evolutions of $\boldsymbol
w(i-1)$, $\alpha$ ($\alpha>1$) is a tuning coefficient that impacts
the update rate and convergence, and $\hat{\sigma}_n^2(i)$ is an
estimate of the noise power, which is assumed to be known at the
receiver. The term $\|\boldsymbol w(i-1)\|^2\hat{\sigma}_n^2(i)$ is
the variance of the inner product of the weight vector with the
noise that provides information on the evolution of $\boldsymbol
w(i-1)$. The time-varying bound provides a smoother evolution of the
weight vector trajectory and thus avoids too high or low values of
the squared norm of $\boldsymbol w$. The proposed SM-CG scheme
utilizes the time-varying bound to create a relation between the
estimated parameters and the environment.

According to $\delta(i)$, we define $\mathcal {H}_i$ to be the set
containing all the estimates of $\boldsymbol w(i)$ for which the
associated array output at time instant $i$ is consistent with the
bound, which is given by $\mathcal H_i=\big\{\boldsymbol w(i)\in\mathbb
C^{m\times1}: |y(i)|^2\leq\delta^2(i)\big\}$. The set $\mathcal H_i$ is
referred to as the \textit{constraint set}, and its boundaries are
hyperplanes. Then, we define the exact \textit{feasibility set}
$\Theta_i$ to be the intersection of the constraint sets over the
instants $l=1, \ldots, i$, which is
\begin{equation}\label{4}
\Theta_i=\mathop{\bigcap_{{l=1}}}_{(s_0, \boldsymbol
r)\in\boldsymbol S}^i\mathcal H_l,
\end{equation}
where $s_0(i)$ is the transmitted data of the desired user and
$\boldsymbol S$ is the set including all possible data pairs
$\{s_0(i),{\boldsymbol r}(i)\}$. It is clear that $\Theta$ should encompass
all the solutions that satisfy the bounded constraint until
$i\rightarrow\infty$. In practice, $\boldsymbol S$ cannot be
traversed all over. Therefore, we define a more practical set
(\textit{membership set}) $\Psi_i=\bigcap_{l=1}^i\mathcal H_l$
instead. The membership set is a limiting set of the feasibility
set. They are equal if the data pairs traverse $\boldsymbol S$
completely.

\subsection{Proposed SM-CG Adaptive Algorithm}

We derive a new adaptive algorithm based on the SM-CG scheme. It
begins with an LCMV optimization problem that incorporates the
constraint on the bound of the array output:
\begin{equation}\label{5}
\begin{split}
&\textrm{minimize}~~\mathbb E[|\boldsymbol w^H\boldsymbol
r(i)|^2]=\boldsymbol w^H\boldsymbol
R\boldsymbol w\\
&\textrm{subject~to}~~\boldsymbol w^H\boldsymbol
a(\theta_0)=\gamma~\textrm{and}~{|y(i)|}^2=\delta^2(i),
\end{split}
\end{equation}
where $\delta(i)$ determines a set of solutions of $\boldsymbol w$
within the constraint set $\mathcal H$ with respect to each time
instant.

The constrained optimization problem can be transformed into an
unconstrained one by the method of Lagrange multipliers. The Lagrangian is
given by
\begin{equation}\label{6}
\begin{split}
&J({\boldsymbol w}(i))=\sum_{l=1}^{i-1}\lambda_1^{i-l}(i)\boldsymbol
w^H(i)\boldsymbol r(l)\boldsymbol r^H(l)\boldsymbol w(i)\\
&+2\lambda_1(i)\mathfrak{R}\big\{|\boldsymbol w^H(i)\boldsymbol
r(i)|^2-\delta^2(i)\big\}+2\lambda_2\mathfrak{R}\big\{\boldsymbol
w^H(i)\boldsymbol a(\theta_0)-\gamma\big\},
\end{split}
\end{equation}
where $\mathfrak{R}\{\cdot\}$ selects the real part of the quantity,
$\lambda_1(i)$ plays the role of the forgetting factor and
Lagrange multiplier with respect to the bounded constraint
and is calculated only if the bounded constraint cannot be satisfied.
The scalar $\lambda_2$ is another Lagrange multiplier to ensure the constraint on the
steering vector of the desired user.

Using the assumption, computing the gradient of $\boldsymbol w(i)$ with
respect to \eqref{6} and equating it to a null vector, we have
\begin{equation}\label{7}
\boldsymbol w(i)=\frac{\gamma\hat{\boldsymbol R}^{-1}(i)\boldsymbol
a(\theta_0)}{\boldsymbol a^H(\theta_0)\hat{\boldsymbol
R}^{-1}(i)\boldsymbol a(\theta_0)},
\end{equation}
where $\hat{\boldsymbol R}(i)=\hat{\boldsymbol
R}(i-1)+\lambda_1(i)\boldsymbol r(i)\boldsymbol r^H(i)$ can be
regarded as an alternative form to estimate the covariance matrix
$\boldsymbol R$. 

To calculate $\boldsymbol w(i)$ efficiently, we adopt the CG-based
adaptive algorithm due to its attractive tradeoff between
performance and complexity. Specifically, we define a CG-based
vector $\boldsymbol v(i)=\hat{\boldsymbol R}^{-1}\boldsymbol
a(\theta_0)$ and use an iterative way to calculate it. The resulting
solution can be written as
\begin{equation}\label{8}
\boldsymbol w(i)=\frac{\gamma\boldsymbol v(i)}{\boldsymbol
a^H(\theta_0)\boldsymbol v(i)},
\end{equation}
where $\boldsymbol v(i)$ is viewed as an intermediate weight vector
for enforcing the constraints and avoiding the matrix inverse. In
the following, we describe a simple CG procedure with only one
iteration per update to calculate $\boldsymbol v(i)$ for computing ${\boldsymbol w}(i)$.

The CG-based vector is expressed by
\begin{equation}\label{9}
\boldsymbol v(i)=\boldsymbol v(i-1)+\alpha(i)\boldsymbol p(i),
\end{equation}
where $\boldsymbol p(i)$ is the direction vector and $\alpha(i)$ is
the corresponding coefficient.

The direction vector $\boldsymbol p(i)$ is obtained by a linear
combination of the previous direction vector and the negative
gradient vector $\boldsymbol g(i)=\boldsymbol
a(\theta_0)-\hat{\boldsymbol R}(i)\boldsymbol v(i)$ of $J(\boldsymbol v(i))=\boldsymbol v^H(i)\hat{\boldsymbol R}(i)\boldsymbol v(i)-2\Re\{\boldsymbol v^H(i)\boldsymbol a(\theta_0)\}$ \cite{Wang},
which is
\begin{equation}\label{10}
\boldsymbol p(i+1)=\boldsymbol g(i)+\beta(i)\boldsymbol p(i),
\end{equation}
where $\beta(i)$ is chosen to provide conjugacy \cite{Chang} for the
direction vectors.

In order to derive the coefficients $\alpha(i)$ and $\beta(i)$, we
consider a recursive form of $\boldsymbol g(i)$
\begin{equation}\label{11}
\boldsymbol g(i)=\boldsymbol g(i-1)-\alpha(i)\hat{\boldsymbol
R}(i)\boldsymbol p(i)-\lambda_1(i)\boldsymbol r(i)\boldsymbol
r^H(i)\boldsymbol v(i-1).
\end{equation}

From \cite{Chang}, the coefficient $\alpha(i)$ should satisfy the
convergence bound $0\leq\boldsymbol p^H(i)\boldsymbol
g(i)\leq 0.5\boldsymbol p^H(i)\boldsymbol g(i-1)$ \cite{Chang}. According to this
bound, premultiplying \eqref{11} with $\boldsymbol p^H(i)$ and
making a rearrangement, we have
\begin{equation}\label{12}
\begin{split}
&\frac{0.5\boldsymbol p^H(i)\boldsymbol
g(i-1)-\lambda_1(i)\boldsymbol p^H(i)\boldsymbol r(i)\boldsymbol
r^H(i)\boldsymbol v(i-1)}{\boldsymbol p^H(i)\hat{\boldsymbol
R}(i)\boldsymbol p(i)}\leq\\
&\alpha(i)\leq\frac{\boldsymbol p^H(i)\boldsymbol
g(i)-\lambda_1(i)\boldsymbol p^H(i)\boldsymbol r(i)\boldsymbol
r^H(i)\boldsymbol v(i-1)}{\boldsymbol p^H(i)\hat{\boldsymbol
R}(i)\boldsymbol p(i)}.
\end{split}
\end{equation}

The relations in \eqref{12} are satisfied if $\alpha(i)$ is
\begin{equation}\label{13}
\alpha(i)=\frac{(1-\eta)\boldsymbol p^H(i)\boldsymbol
g(i-1)-\lambda_1(i)\boldsymbol p^H(i)\boldsymbol r(i)\boldsymbol
r^H(i)\boldsymbol v(i-1)}{\boldsymbol p^H(i)\hat{\boldsymbol
R}(i)\boldsymbol p(i)},
\end{equation}
where $0\leq\eta\leq0.5$.

For $\beta(i)$, since $\boldsymbol p^H(i)\hat{\boldsymbol
R}(i)\boldsymbol p(i+1)=0$, it follows that
\begin{equation}\label{14}
\beta(i)=-\frac{\boldsymbol p^H(i)\hat{\boldsymbol R}(i)\boldsymbol
g(i)}{\boldsymbol p^H(i)\hat{\boldsymbol R}(i)\boldsymbol p(i)}.
\end{equation}

The coefficient $\lambda_1(i)$ is important to obtain the filter
parameters. The SM technique provides an adaptive strategy to obtain
it following the changes of the scenarios. Substituting \eqref{9}
and \eqref{13} into the bounded constraint in \eqref{5} and performing some algebraic manipulations, we obtain
\begin{equation}\label{15}
\lambda_1(i)=\frac{\lambda_{11}(i)-\lambda_{12}(i)}{\lambda_{13}(i)-\lambda_{14}(i)},
\end{equation}
where\\
{\small $\lambda_{11}(i)=\tau_1(i)\textrm{sign}\{\tau_1(i)-\tau_2(i)\}$; \\ $\lambda_{12}(i)=\tau_3(i)\textrm{sign}\{\tau_3(i)-\tau_4(i)\}$;\\
$\lambda_{13}(i)=\tau_2(i)\textrm{sign}\{\tau_1(i)-\tau_2(i)\}$;\\
$\lambda_{14}(i)=\tau_4(i)\textrm{sign}\{\tau_3(i)-\tau_4(i)\}$\\
$\tau_1(i)=\delta(i)\boldsymbol v^H(i-1)\boldsymbol
a(\theta_0)\boldsymbol p^H(i)\hat{\boldsymbol R}(i)\boldsymbol
p(i)+\delta(i)(1-\eta)\boldsymbol g^H(i-1)\boldsymbol
p(i)\boldsymbol p^H(i)\boldsymbol a(\theta_0)$\\
$\tau_2(i)=\boldsymbol v^H(i-1)\boldsymbol r(i)\boldsymbol
r^H(i)\boldsymbol
p(i)\boldsymbol p^H(i)\boldsymbol a(\theta_0)$\\
$\tau_3(i)=\boldsymbol v^H(i-1)\boldsymbol r(i)\boldsymbol
p^H(i)\hat{\boldsymbol R}(i)\boldsymbol p(i)+(1-\eta)\boldsymbol
g^H(i-1)\boldsymbol p(i)\boldsymbol p^H(i)\boldsymbol r(i)$\\
$\tau_4(i)=\boldsymbol v^H(i-1)\boldsymbol r(i)\boldsymbol
r^H(i)\boldsymbol p(i)\boldsymbol p^H(i)\boldsymbol r(i)$}.

The proposed SM-CG algorithm is summarized in Table \ref{tab:CG-SM},
where the initialization is given to ensure the constraint on the
steering vector of the desired user and to start the update. From
Table \ref{tab:CG-SM}, the coefficient $\lambda_1(i)$ is calculated
only if the bounded constraint cannot be satisfied, so as the update
procedure. The data-selective updates save the computational cost
significantly. Compared with most existing CG-based algorithms
\cite{Wang}, the estimation of $\boldsymbol v(i)$ in the proposed
algorithm only runs one iteration per update, which further reduces
the complexity. All the estimates $\boldsymbol w(i)$ ensuring the
bounded constraint until time instant $i$ are in the feasibility set
$\Theta_i$.

\begin{table}[!t]
\centering
    \caption{THE PROPOSED SM-CG ALGORITHM}     
    \label{tab:CG-SM}
    \begin{small}
        \begin{tabular}{|l|}
\hline
\bfseries {Initialization:}\\
\hline
~~~~~$\boldsymbol g(0)=\boldsymbol p(1)=\boldsymbol a(\theta_0)$;~~$\boldsymbol w(0)=\boldsymbol a(\theta_0)/\|\boldsymbol a(\theta_0)\|^2$.\\
\hline
\bfseries {For each time instant} $i=1,\ldots, N$\\
\hline
~~~~~$y(i)={\boldsymbol w}^H(i-1){\boldsymbol r}(i)$\\
~~~~~$\delta(i)=\beta\delta(i-1)+(1-\beta)\sqrt{\alpha\|\boldsymbol
w(i-1)\|^2\hat{\sigma}_n^2(i)}$\\
~~~~~\bfseries {if} ~~~$|y(i)|^2\geq\delta^2(i)$\\
~~~~~~~~~~~$\lambda_1(i)=\frac{\lambda_{11}(i)-\lambda_{12}(i)}{\lambda_{13}(i)-\lambda_{14}(i)}$\\
~~~~~~~~~~~$\hat{\boldsymbol R}(i)=\hat{\boldsymbol
R}(i-1)+\lambda_1(i)\boldsymbol r(i)\boldsymbol r^H(i)$\\
~~~~~~~~~~~$\alpha(i)=\frac{(1-\eta)\boldsymbol p^H(i)\boldsymbol
g(i-1)-\lambda_1(i)\boldsymbol p^H(i)\boldsymbol r(i)\boldsymbol
r^H(i)\boldsymbol v(i-1)}{\boldsymbol p^H(i)\hat{\boldsymbol
R}(i)\boldsymbol p(i)}$\\
~~~~~~~~~~~$\boldsymbol v(i)=\boldsymbol v(i-1)+\alpha(i)\boldsymbol p(i)$\\
~~~~~~~~~~~$\boldsymbol g(i)=\boldsymbol
g(i-1)-\alpha(i)\hat{\boldsymbol R}(i)\boldsymbol
p(i)$\\
~~~~~~~~~~~~~~~~~~~~~~$-\lambda_1(i)\boldsymbol r(i)\boldsymbol
r^H(i)\boldsymbol v(i-1)$\\
~~~~~~~~~~~$\beta(i)=-\frac{\boldsymbol p^H(i)\hat{\boldsymbol
R}(i)\boldsymbol g(i)}{\boldsymbol p^H(i)\hat{\boldsymbol R}(i)\boldsymbol p(i)}$\\
~~~~~~~~~~~$\boldsymbol p(i+1)=\boldsymbol g(i)+\beta(i)\boldsymbol p(i)$\\
~~~~\bfseries {else}\\
~~~~~~~~~~~${\boldsymbol v}(i)={\boldsymbol v}(i-1)$\\
~~~~\bfseries {end}\\
~~~~$\boldsymbol w(i)=\frac{\gamma\boldsymbol v(i)}{\boldsymbol
a^H(\theta_0)\boldsymbol v(i)}$\\
\hline
    \end{tabular}
    \end{small}
\end{table}

Regarding the complexity, the algorithms with the SM technique
require much less computational cost than their counterparts without
the SM technique due to the data selective updates. Since the
calculations of the array output $y(i)$ and the time-varying bound
$\delta(i)$ are the same for the SM-type algorithms, we check their
procedures during the updates to compare the complexity. The
proposed algorithm needs around $2\tau N m^2$ additions and $2\tau N
m^2$ multiplications for the operation, where $\tau$ ($0<\tau\leq1$)
is the update rate. These computational requirements are greater
than those of the SG-based algorithm \cite{Diniz},
\cite{Lamare} ($3\tau N m$ for additions and $4\tau N m$ for
multiplications) but much less than the RLS-based algorithm
\cite{Huang}, \cite{Lamare} ($4\tau N m^2$ for additions and $5\tau
N m^2$ for multiplications), and the AP-based algorithm
\cite{Diniz2} ($\tau N m L^2$ for additions and multiplications)
with $L$ being the size of the signal matrix. In the following part,
we will see that the proposed algorithm spends less updates than the
existing algorithms but has a fast convergence and
shows an excellent tracking performance.

\section{Simulations Results}

We evaluate the output signal-to-interference-plus-noise ratio
(SINR) performance of the proposed and existing algorithms for the
LCMV beamformer. Specifically, we compare the proposed algorithm
with the SG and RLS algorithms with/without the SM technique
\cite{Haykin}, \cite{Chang}, \cite{Nagaraj2}, \cite{Lamare}, and the
AP algorithm with the SM technique (SM-AP) \cite{Diniz2}. We assume
that there is one desired user in the system and the related DOA is
known beforehand by the receiver. The results are averaged by $500$
runs. We consider the binary phase shift keying (BPSK) modulation
scheme and set $\gamma=1$ for the algorithms. Simulations are
performed with a ULA containing $m=16$ sensor elements with
half-wavelength interelement spacing.

\begin{figure}[h]
    \centerline{\psfig{figure=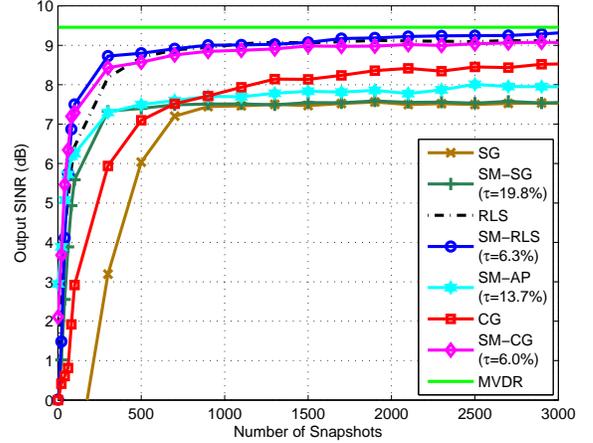,height=64.54mm} }
    \caption{Output SINR versus the
number of snapshots.}
    \label{fig:cg_sm_sinr}
\end{figure}

In the first experiment, there are $q=10$ users in the system. The
input signal-to-noise ratio (SNR) is $10$ dB and the
interference-to-noise ratio (INR) is $30$ dB. We set $\alpha=21$,
$\beta=0.9$ and $\eta=0.5$ for the proposed algorithm. Note that
$\lambda_1(i)$ should be a small positive value close to but less
than $1$ in accordance with the setting of the forgetting factor. In simulations, we limit its range
$0.1\leq\lambda_1(i)\leq0.999$. In Fig. \ref{fig:cg_sm_sinr}, the
curves of all the algorithms converge to their steady-state
following the increase of the snapshots. The algorithms with the SM
technique show faster convergence rates than the standard algorithms. The
proposed algorithm has a convergence comparable to that of the
SM-RLS algorithm and the steady-state performance has a SINR level
close to that of the MVDR solution. It only requires $\tau=6.0\%$ updates
($178$ updates for $3000$ snapshots) for the filter design, which is
lower than those of the existing algorithms and thus reduces the
complexity significantly.

\begin{figure}[h]
    \centerline{\psfig{figure=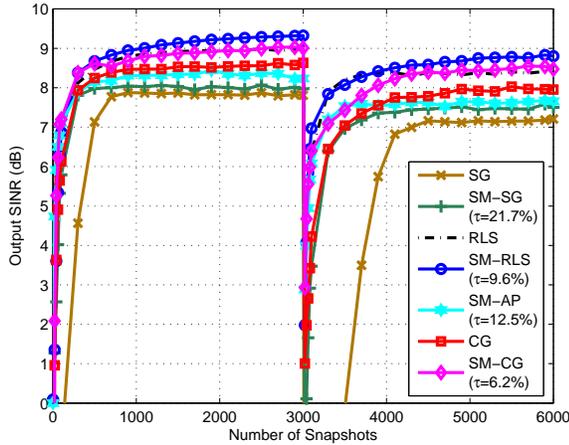,height=64.54mm} }
    \caption{Output SINR versus the
number of snapshots in dynamic scenario with additional users enter
and/or leave the system.}
    \label{fig:cg_sm_more}
\end{figure}

The next experiment shows the output SINR performance for the
proposed algorithm under a non-stationary scenario, namely, when the
number of users changes in the system. The system starts with $q=8$
users including one desired user. The input SNR is $10$ dB and the
INR is $35$ dB. The coefficients are the same as those in Fig.
\ref{fig:cg_sm_sinr} except $\alpha=23$. From the first stage (first
$3000$ snapshots) of Fig. \ref{fig:cg_sm_more}, the proposed
algorithm converges quickly to the steady-state. The scenario
experiences a sudden change at $N=3000$. We have $4$ more
interferers entering the system, which results in the performance
degradation for the studied algorithms. The algorithms with the SM
technique track this change rapidly and reach the steady-state since
the data-selective updates reduce the number of updates and keep a
faster convergence rate. Besides, the time-varying bound provides
information for them to follow the changes of the scenario. The
change also influences the update rate of the algorithms. According
to the statistics, the update rate of the proposed algorithm
($\tau=6.2\%$) is rather insensitive to the change and saves
computational cost.

\section{Conclusion}
In this paper, we have introduced a
new adaptive filtering strategy that combines the SM
technique with the adaptive CG algorithm for the LCMV beamformer design.
We defined an LCMV optimization problem related to a
constraint on the bound of the array output and proposed a CG-based
adaptive algorithm for implementation. The proposed algorithm
performs the data-selective updates to obtain the filter
parameters. For the update, a CG-based vector has been devised to create
a relation between the covariance matrix inverse and the steering
vector of the desired user. The proposed SM-CG algorithm calculates the
CG-based vector to encompass a space of feasible solutions with
respect to each time instant and to enforce the constraints. The
proposed algorithm exhibits a very good convergence and tracking
performance with relatively low computational cost.

\end{document}